\newcommand{\be}{\begin{equation}}
\newcommand{\ee}{\end{equation}}
\newcommand{\bea}{\begin{eqnarray}}
\newcommand{\eea}{\end{eqnarray}}
\newcommand{\bean}{\begin{eqnarray*}}

\newcommand{\eean}{\end{eqnarray*}}
\font\upright=cmu10 scaled\magstep1
\font\sans=cmss10
\newcommand{\ssf}{\sans}
\newcommand{\stroke}{\vrule height8pt width0.4pt depth-0.1pt}
\newcommand{\Z}{\hbox{\upright\rlap{\ssf Z}\kern 2.7pt {\ssf Z}}}

\newcommand{\C}{{\rlap{\rlap{C}\kern 3.8pt\stroke}\phantom{C}}}
\newcommand{\R}{\hbox{\upright\rlap{I}\kern 1.7pt R}}
\newcommand{\CP}{\C{\upright\rlap{I}\kern 1.5pt  P}}
\newcommand{\PP}{\hbox{\upright\rlap{I}\kern 1.5pt  P}}

\newcommand{\identity}{{\upright\rlap{1}\kern 2.0pt 1}}

\newcommand{\HH}{\mbox{\hbox{\upright\rlap{I}\kern 1.7pt H}}}

\newcommand{\fr}{\frac}

\newcommand{\ra}{\rightarrow}

\newcommand{\pr}{\partial}
\newcommand{\hs}{\hspace{5mm}}

\newcommand{\ve}{\varepsilon}

\input{epsf}

\documentstyle[12pt,a4wide]{article}

\begin{document}
\title{\vskip -70pt
\begin{flushright}
\end{flushright}\vskip 50pt
{\bf \large \bf
Soliton Dynamics in a 2D Lattice Model with 
Nonlinear Interactions}}
\author{T. Ioannidou$^1$\thanks{P{\it ermanent Address:} Institute of
Mathematics, University of Kent, Canterbury CT2 7NF, UK}\,,
J. Pouget$^1$ and  E. Aifantis$^2$\thanks{Center for Mechanics of
Materials and Instabilities, Michigan Technological University, Houghton, MI
 49931, USA}
\\[10pt]
$^1${\normalsize  {\sl Laboratoire de Mod\'elisation en 
M\'ecanique (associ\'e au CNRS),}}\\
{\normalsize  {\sl Universit\'e Pierre et Marie Curie, 
Tour 66, 4 place Jussieu,}}\\
{\normalsize  {\sl 75252 Paris C\'edex 05, France}}\\[10pt]
$^2${\normalsize  {\sl Laboratory of Mechanics and Materials, 
Polytechnic School,}}\\
{\normalsize  {\sl Aristotle University of Thessaloniki, 
54006, Thessaloniki, Greece}}} \date{}
\maketitle

\begin{abstract}
This paper is concerned with a lattice  model 
which is suited to square-rectangle transformations
characterized by two strain components.
The microscopic model involves nonlinear and competing
interactions, which play a key role in the stability of soliton solutions
and emerge from interactions as a function of particle pairs 
and noncentral type or bending forces.
Special attention is devoted to the continuum approximation
of the two-dimensional discrete system with the view
of including the leading discreteness effects at the 
continuum description.
The long time evolution of the localized structures is governed
by an asymptotic integrable equation of the  
Kadomtsev-Petviashvili I type which allows the explicit 
construction of moving multi-solitons on the lattice.
Numerical simulation performed at the discrete system 
investigate the stability and dynamics  of multi-soliton in the lattice 
space.

\end{abstract}
\section{Introduction}

A lot of interest has recently been devoted to spatio-temporal patterns
 as well as the associated defects and dynamics such as standing-wave
patterns, localized structures including solitons or oscillating patterns.
These structures become fundamental in the study of phase transitions 
which are usually accompanied by the appearance of defects: dislocation 
motions, grain boundaries, domain wall structures and twinnings
\cite{GR}-\cite{PDE}.
One of the aims of the present research is to understand how solitons arising
at the microscale  (ie at the level of the lattice model), are
able to organize the system at the macroscale and what the selection
properties of the nonlinear structures are.
The present work is particularly motivated by the existence of solitonic
structures  occurring in phase transformations in crystalline alloys.

The same lattice model has been successfully used to examine 
the formation  of localized strain patterns decaying in all
directions \cite{P1}-\cite{P3}.
Then the corresponding research results in the partial softening of
 the transverse-acoustic phonon branch at a nonzero wave number;
the positive curvature of the dispersion branch at the
long wavelength limit; and the shearing motion of the atomic 
planes along the stacking direction leading to spatially
arranged structures made of martensitic twin bands and the
existence of strain solitary waves describing the coherent
movement of martensitic domains.
More precisely, the nonlinear dynamics of the two-dimensional model allows
us to examine the stability of the lattice using numerical simulations
which have shown the formation of localized strain structures 
emerging from an instability mechanism \cite{P1,P3}.

In the present paper we continue the investigation of the
properties of the two-dimensional lattice model in order
to study the existence, stability and dynamics 
of lattice multi-soliton configurations.
In fact, the study provides the critical values of the coefficients
of the lattice model for which lattice solitons exist and
move along the plane, using the fact that the continuum limit of 
 the long time evolution of the patterns is governed by an asymptotic
equation of the Kadomstev-Petsviashili I type.
Recently in \cite{Pot}, the  Kadomstev-Petsviashili equation 
has been derived for quasiplane waves by considering oscillations
of a two-dimensional square lattice array of atoms  with interactions
 only between the nearest neighbors.

This paper is organized as follows: in section 2 we introduce the 
lattice model,  while in section 3 we deal with its continuum approximation
and show that it leads to the  Kadomstev-Petsviashili I equation
for long time evolution.
Finally, in section 4 numerical simulations of the discrete model based
on the Kadomstev-Petsviashili I multi-solitons are performed  and their dynamics 
are discussed in some detail.

\section{Lattice Model}

\begin{figure}
\vskip -2cm
\hskip 4cm
\epsfxsize=8cm\epsfysize=8cm\epsffile{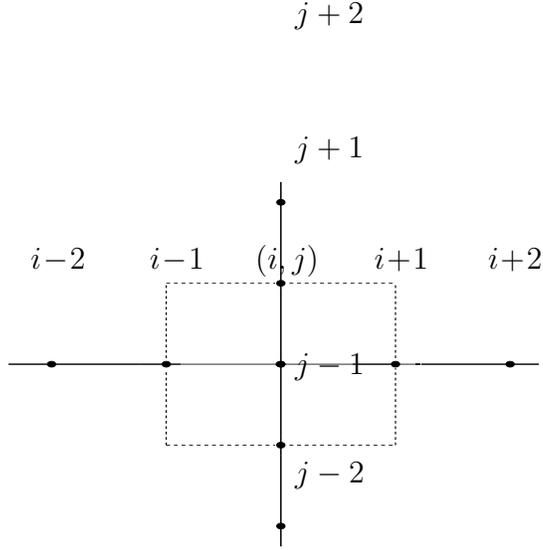}
\hfill \hspace{-3mm}
\put(-225,243){$j+2$}
\put(-225,192){$j+1$}
\put(-225,110){$j-1$}
\put(-225,69){$j-2$}
\put(-240,150){$(i,j)$}
\put(-195,150){$i\!+\!1$}
\put(-152,150){$i\!+\!2$}
\put(-280,150){$i\!-\!1$}
\put(-325,150){$i\!-\!2$}
\vskip -2cm
\caption{Two-dimensional lattice model with the detail of interatomic interactions
 between the first and second neighbors in the $i,j$ direction.}
\label{fig-lat}
\end{figure}

Let us consider an atomic plane made of squares parallel to the $i$ and 
$j$ directions presented in Figure \ref{fig-lat}.
Such a lattice model can be extracted from the cubic lattice of crystalline
alloys (like the fcc symmetry of In-Tl, Fe-Pd and other crystals).
The model describes a cubic tetragonal transformation of the lattice.
A particle of the lattice plane is located by $(i,j)$.
After deformation of the lattice, the particle undergoes a displacement
defined by $u_{i,j}=u(i,j)$ along the $i$ and $j$ directions.

The particles interact via two kinds of interatomic
potentials: (i) interactions between first-nearest neighbors considered as
functions of the particle pairs in the $i$ and $j$ diagonal directions
and (ii) interactions involving noncentral forces or three body interactions
between the first-nearest and second-nearest neighbors in the $i$ and $j$ 
directions.
The potential describing the first interactions possesses stable, unstable 
or metastable states accordingly to a control parameter which is connected
with temperature \cite{KG}.
The latter interactions amount to describing, at the microscopic level,
the resistance of the crystalline cell to twisting and bending \cite{P1,P2}
 and thus provide competing interactions \cite{Kr1,Kr2}.
The noncentral interactions are of particular interest for the competing
interactions and stability of the nonlinear structures.
On using the invariance of the lattice energy under translations and 
rotations, the following potential functional was introduced in \cite{P3}
\bea
{\cal V}\!\!\!\!&=&\!\!\!\!\sum_{(i,j)}\biggl[\Phi\, S_{i,j}
+\fr{\beta}{2}\,G_{i,j}^2+\fr{\delta}{2}
\left\{(\Delta_L^+S_{i,j})^2+(\Delta_T^+G_{i,j})^2\right\}\nonumber\\
&&\,\,\,\,\,\,\,+\fr{\eta}{2}\left(\left\{\Delta_L^+\left[S_{i+1,j}
+2S_{i,j}+S_{i-1,j}\right]\right\}^2
+\left\{\Delta_T^+\left[G_{i+1,j}+2G_{i,j}+G_{i-1,j}\right]
\right\}^2 \right) \biggl]
\label{len}
\eea
where the discrete deformations are defined as
\bea
S_{i,j}=u_{i,j}-u_{i-1,j}, \hs \hs \hs G_{i,j}=u_{i,j}-u_{i,j-1}
\label{SG}
\eea
and the potential $\Phi$ is given by
\be
\Phi(S_{i,j})=\fr{1}{2}\,\alpha_1 \,S_{i,j}^2-\fr{1}{3}\,\alpha_2 \,S_{i,j}^3.
\label{poten}
\ee
For simplicity, the lattice energy (\ref{len}) has been chosen to be 
dimensionless.
The first and second terms in (\ref{len}) represent the nonlinear and linear
potentials coming from the particle pair interactions where 
($\alpha_1$, $\alpha_2$) and $\beta$ are the lattice force coefficients for the
longitudinal and shear deformations, respectively.
The third and fourth parts of (\ref{len}) hold for the actions of the 
noncentral interactions in the $i$ and $j$ directions.
The interactions are characterized by the parameters $\delta$ and $\eta$
for the actions between first- and second-nearest particles, respectively.
On the other hand, the operators $\Delta_L^+$ and $\Delta_T^+$ hold for 
the forward first-order finite difference in the $i$ and $j$ directions:
$\Delta_L^+S_{i,j}=S_{i+1,j}-S_{i,j}$ and $\Delta_T^+G_{i,j}=G_{i,j+1}-G_{i,j}$.
{\it Remark:} Due to (\ref{SG}) the noncentral interactions in (\ref{len})
are of the form $u_{i-1,j}+u_{i+1,j}-2u_{i,j}$ for the first-nearest neighbor
interaction and of the form  $u_{i-2,j}+u_{i+2,j}-2u_{i,j}$ 
for the second-nearest neighbor interaction (same for the $G_{i,j}$ 
deformation).

In addition, a one-dimensional version of the model can be obtained,
reduced from the complete two-dimensional one, in which
the shearing motion of the atomic planes along the stacking direction
is modeled by arrays of martensitic and austenitic solitary waves.
This reduced system has been examined in detail,
 and it has been observed that localized structures emerge of 
arrays of elastic solitary waves \cite{P1,P2}.

By introducing the kinetic energy associated with the displacement 
$u_{i,j}$ (for unit mass) as
\be
{\cal T}=\sum_{(i,j)}\fr{1}{2}\, \dot{u}_{i,j}
\ee
the  corresponding difference-differential  equations of 
motion for $u_{i,j}$, deduced from the Hamiltonian 
$H=\cal{T}+{\cal V}$, are given by
\be
\ddot{u}_{i,j}=\Delta_L^+ \Sigma_{L_{i,j}}+\Delta_T^+ \Sigma_{T_{i,j}}
\label{glatt}
\ee
where the discrete stresses are defined as
\bea
\Sigma_{L_{i,j}}&=&\sigma_{i,j}-\Delta_L^- \,\chi_{L_{i,j}}\label{s1}\\
\Sigma_{T_{i,j}}&=&\beta\, G_{i,j}-\Delta_T^- \,\chi_{T_{i,j}}\label{s2}\\
\sigma_{i,j}&=&\alpha_1 \,S_{i,j}-\alpha_2\, S_{i,j}^2\label{str}\\
\chi_{L_{i,j}}&=&\Delta_L^+\left\{\delta \,S_{i,j}+\eta \left[S_{i+2,j}+
4S_{i+1,j}+6S_{i,j}+4S_{i-1,j}+S_{i-2,j}\right] \right\}\label{chi1}\\
\chi_{T_{i,j}}&=&\Delta_T^+\left\{\delta \,G_{i,j}+\eta \left[G_{i,j+2}+
4G_{i,j+1}+6G_{i,j}+4G_{i,j-1}+G_{i,j-2}\right] \right\}\label{chi2}.
\eea
Equations (\ref{s1}-\ref{s2}) correspond to the discrete macroscopic 
stresses due to the fact that the deformations $S_{i,j}$ and $G_{i,j}$ are 
functions of the discrete displacements $u_{i,j}$.
Note that the stress (\ref{str}) which follows from the potential $\Phi$ since
$\sigma_{i,j}=\pr \Phi/\pr S_{i,j}$ is nonlinear in terms of the strain 
$S_{i,j}$.
Also, the microscopic stresses (\ref{chi1}-\ref{chi2}) emerging from 
the noncentral forces are functions of the discrete variations
of the deformations  $S_{i,j}$ and $G_{i,j}$ in the $i$ and $j$ direction,
respectively.

Due to the strongly nonlinear nature of the problem,
these equations are not managable except for the linear 
problem which has been examined in Ref. \cite{P4}.
In this paper, these equations are solved using numerical simulations 
with appropriate initial and boundary conditions.
More precisely, using the continuum approximation a
quasi-continuum model has been obtained which includes the
leading discreteness effects and allows us to investigate 
the existence of soliton configurations and study their dynamics.

\section{Continuum Approximation}

In order to describe the lattice dynamics at the quasi-continuum scale
we assume that the discrete functions are slowly varying over 
the lattice spacing.
In fact, both the deformations (\ref{SG}) have been expanded using Taylor's 
series up to third order, that is 
 $S=hu_x-\fr{h^2}{2}u_{xx}+\fr{h^3}{6}u_{xxx}$ (where $h$ is 
the lattice spacing);
while all other terms have been expanded for long wavelength  up to
fourth order.
After some classical algebras the continuum equations of motion for the
 deformation $u(x,y,t)$ are
\be
u_{tt}=\alpha_1\,u_{xx}-\alpha_2\,(u^2)_{xx}+\delta_L\, u_{xxxx}+\delta_T\,
u_{yyyy}+\beta u_{yy}\\
\label{cont}
\ee
for  dimensionless space unit, ie rescaling  $x \ra x/h$ and $y \ra y/h$.
The coefficients $\delta_L$ and $\delta_T$ are given in terms of 
the coefficients of the model, ie
\be
\delta_L=\fr{\alpha_1}{12}-\delta-16\eta,\hs\hs \hs
\delta_T=\fr{\beta}{12}-\delta-16\eta.
\label{del}
\ee

\subsection{Asymptotic Model}
In order to understand the evolution of the localized structures over
a large scale of time of the order of $\varepsilon^{-1}$ and for a weakly 
nonlinear medium, we consider
the asymptotic equation derived from the continuum equation (\ref{cont}).
By assuming that the contribution of the nonlinear term is 
weak throughout we rescale the nonlinear and dispersive terms
by introducing a small parameter $\varepsilon<1$ as follows
\be
\alpha_2=\varepsilon \bar{\alpha}, \hs \hs \hs 
\delta_L=\varepsilon \bar{\delta}_L,
\hs \hs \hs  \delta_T=\varepsilon \bar{\delta}_T.
\label{epar}
\ee
Then for large times (of order $\varepsilon^{-1}$), the asymptotic expansion 
of the displacement
field is
\be
u(x,y,t)=u_0(\xi,Y,\tau)+\ve u_1(x,y,t)+O(\ve^2)
\label{aex}
\ee
where $\xi=x-ct$ is the stretch phase variable, $Y=\ve^{1/2}y$ is
the stretch transverse variable and $\tau=\ve t$ is the slow time
variable.

Using equations (\ref{epar}) and (\ref{aex}) and keeping terms of
order $\ve$ only, equation (\ref{cont}) becomes
\be
2cu_{0_{\tau\xi}}-\bar{\alpha} \,(u^2_{0_{\xi}})_{\xi}+\bar{\delta}_L\,
u_{0_{\xi\xi\xi\xi}}+\beta\, u_{0_{YY}}=u_{1_{tt}}-\alpha_1 u_{1_{xx}}
-\beta u_{1_{yy}}
\label{all}
\ee
where we have set  the sound velocity in the $x$  direction to be equal 
to  $c=\sqrt{\alpha_1}$ (for $\alpha_1>0$).

Then the secularity condition \cite{Nay} implies that both
the left- and right-hand side of (\ref{all}) have to equal zero.
Thus, the right-hand side leads to the standard two-dimensional
linear wave equation for $u_1$, while the left-hand side gives
the long time evolution of $u_0$ defined by
\be
\left(u_{0_{T}}+\bar{\alpha} (u^2_{0})_{\xi}+\tilde{\delta}_L
 u_{0_{\xi\xi\xi}} \right)_\xi=\beta \,u_{0_{YY}}.
\label{KPa}
\ee
Note that the following change of variables and of parameters have been
considered $T=-\tau/2c$ and $\tilde{\delta}_L=-\bar{\delta}_L$ in 
order that (\ref{KPa}) transforms into a standard form, and that
for specific choices of the parameters $\bar{\alpha}$, 
$\tilde{\delta}_L$ and $\beta>0$ equation (\ref{KPa}) becomes the  
Kadomtsev-Petviashvili I equation \cite{KP} 
(see later). Note that equation (\ref{KPa}) was first derived in 
\cite{P3} using Fourier images.

\subsection{The Kadomstev-Petsviashili Equations}

Nonlinear quasi-one dimensional waves (with $y$ much larger that $x$)
in a weakly dispersive medium are described by the dimensionless
 Kadomstev-Petsviashili (KP) equations
\be
\left(u_t+3 (u^2)_x+ u_{xxx} \right)_x= \pm \,u_{yy}
\label{KP}
\ee
where the $\pm$ on the right-hand side of (\ref{KP}) is determined
by the dispersive property of the system; ie the upper sign is usually
refered to as positive dispersion and the corresponding equation is known 
as KPI (which is the case we study here).

The exact multi-soliton solutions of KPI can be constructed by
different methods (see, for example, Refs. \cite{MZ,SA}).
For our purposes it is convenient to write them in the Hirota form, ie
\be
u(x,y,t)=2\,\fr{\pr^2}{\pr x^2}\ln \phi
\label{sol}
\ee
where $\phi$ is the determinant of a $2n\times 2n$ matrix 
(for $n$ number of solitons) given by
\be
\phi=\det \left[\left(x+p_ky+p_k^2t+\theta_k\right)\delta_{kl}
+\left(1-\delta_{kl}\right) \fr{2 \,\sqrt{3}\,i}{p_k-p_l}\right].
\label{sol1}
\ee
Here $\delta_{kl}$ is the Kronecker symbol, the indices take values 
$k,l=1,2,\cdots,2n$, while the complex constants $p_k$ and $\theta_k$ 
determine the velocity and the phase of each soliton, where
$p_{k+n}=\bar{p}_k$ and $\theta_{k+n}=\bar{\theta}_k$.

The one soliton solution given by (\ref{sol}-\ref{sol1}), when
$p_1=i$ and $\theta_1=60$, has the simple expression
\be
u(x,y,t)=4\fr{-\left(x-t+60\right)^2+y^2+3}
{\left[\left(x-t+60\right)^2+y^2+3\right]^2}.
\label{1so}
\ee
The corresponding configuration consists of a soliton 
which for $t=0$ is situated at 
the points $(x,y)=(-60,0)$  while for $t\neq 0$ travels 
along the $x$-axes without changing its shape and 
 with   constant velocities: 
$(v_x,v_y)=\left(|p_1|^2, -2\,\Re(p_1)\right)$.

\begin{figure}
\vskip -3cm
\hskip 2cm
\put(20,190){$t=-4$} 
\epsfxsize=8cm\epsfysize=8cm\epsffile{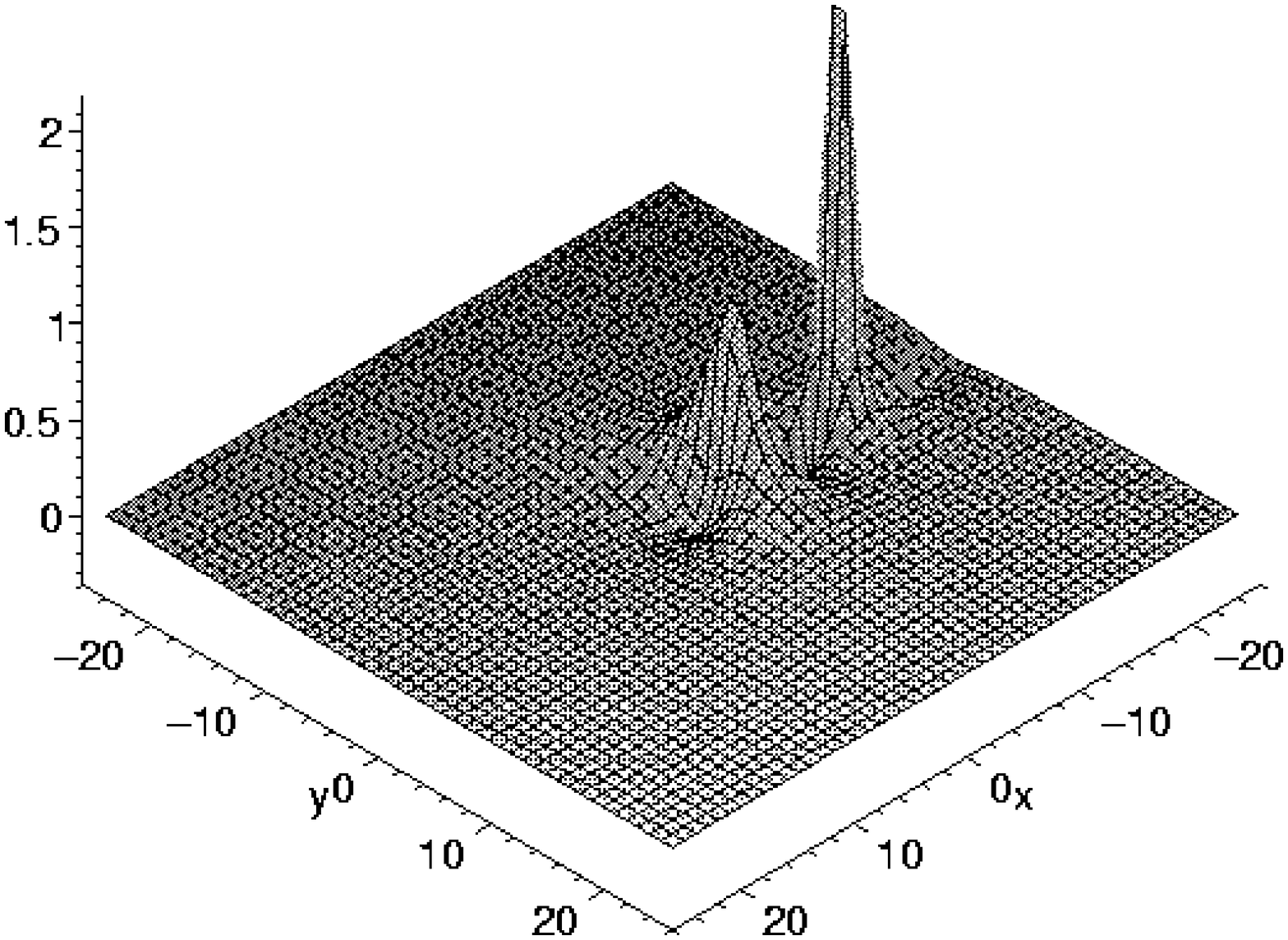}
\hfill 
\put(-30,190){$t=4$}
\vskip -9cm
\hskip 8cm
\epsfxsize=8cm\epsfysize=8cm\epsffile{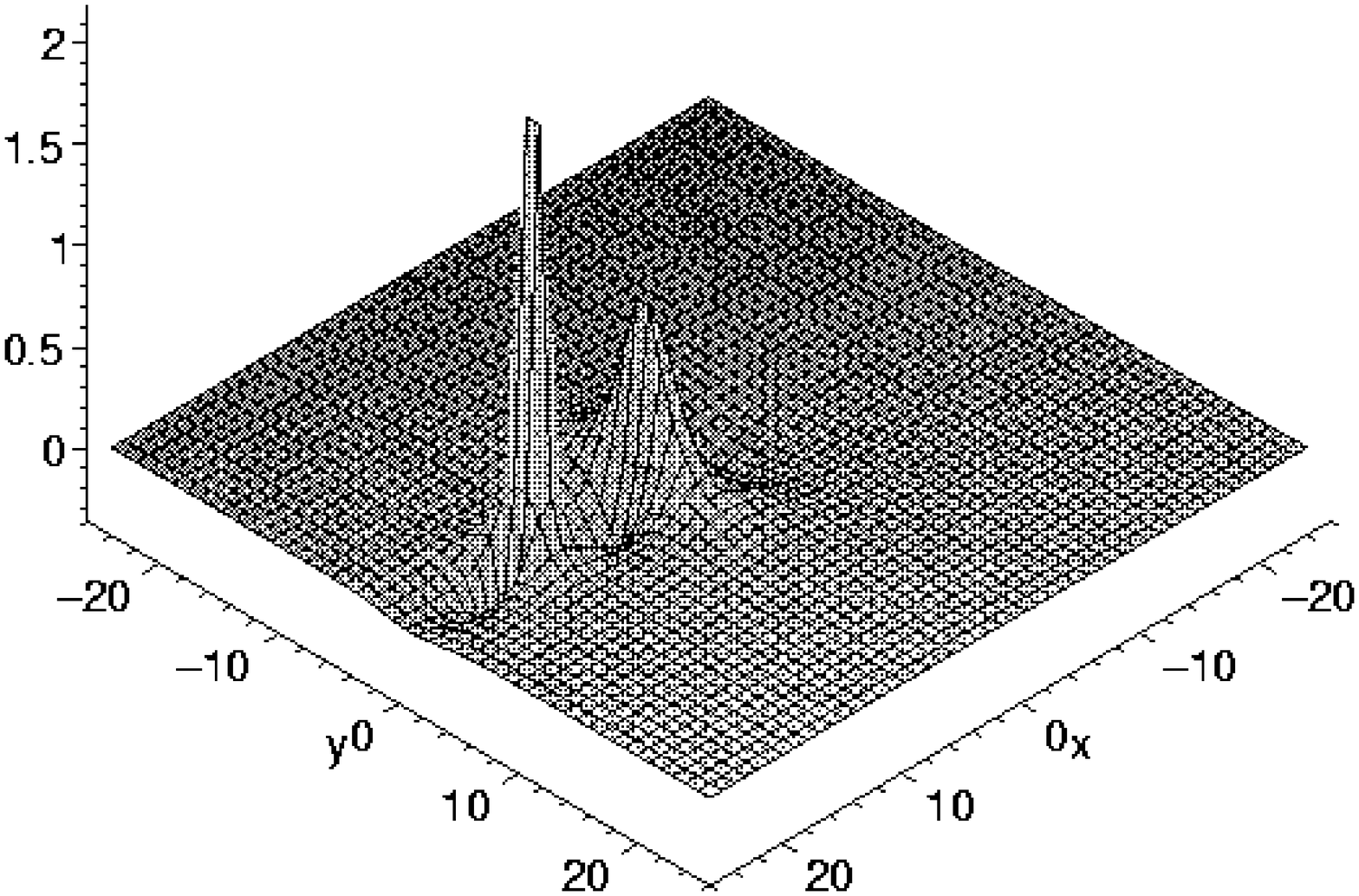}
\caption{The  two soliton solution (\ref{sol}-\ref{sol1}) of the KPI equation 
at different times with trivial  scattering behaviour. 
The taller soliton passes through the smaller one with no
phase shift or radiation.}
\label{fig-2s}
\end{figure}   

In addition, the corresponding multi-soliton solutions
 (\ref{sol}-\ref{sol1})
are asymptotically free at $t\ra \pm \infty$  and move along their 
unperturbed trajectories without changing their shapes
and initial parameters after collision, ie they scatter trivially.
Figure \ref{fig-2s} represents snapshots of the two soliton solution 
(\ref{sol}-\ref{sol1}) for  $p_1=i$, $p_2=2i$ and $\theta_1=\theta_2=0$.

Note that in the framework of plasma physics the classical KP
 equations suffer transverse instabilities accordingly to the sign 
of the dispersive terms as well as possessing soliton solutions
\cite{LS}.

\section{Numerical Simulations}

In this section, we investigate numerically the existence of
lattice multi-solitons using the fact that the lattice model
asymptotically leads to an integrable equation.
We use a numerical scheme by directly considering the lattice
equations for the displacement $u_{i,j}$ 
given by 
\bea
\ddot{u}_{i,j}\!\!\!\!&=&\!\!\!\!\alpha_1\!\left(u_{i+1,j}+u_{i-1,j}
\!-\!2u_{i,j}\right)\!+\!\alpha_2\!\left(2u_{i,j}\!-\!u_{i-1,j}
\!-\!u_{i+1,j}\right)\!\left(u_{i+1,j}\!-\!u_{i-1,j}\right)\!+\!
\beta\!\left(u_{i,j+1}\!+\!u_{i,j-1}\!-\!2u_{i,j}\right)\nonumber\\
&-&\!\!\!\!\delta\left(u_{i+2,j}-4u_{i+1,j}+6u_{i,j}-4u_{i-1,j}
+u_{i-2,j}\right)
-\delta\left(u_{i,j+2}-4u_{i,j+1}+6u_{i,j}-4u_{i,j-1}
+u_{i,j-2}\right)\nonumber\\
&-& \!\!\!\!\eta\left(u_{i+4,j}-4u_{i+2,j}+6u_{i,j}-4u_{i-2,j}
+u_{i-4,j}\right) 
-\eta\left(u_{i,j+4}-4u_{i,j+2}+6u_{i,j}-4u_{i,j-2}
+u_{i,j-4}\right)\nonumber\\
\label{latt}
\eea
which follow from (\ref{SG}) and (\ref{glatt}).
The numerical simulations are performed by employing
 a Runge-Kutta method of fourth order and by imposing
 periodic boundary conditions, ie
\be
u_{i+N,j}=u_{i,j}, \hs \hs \hs u_{i,j+N}=u_{i,j}
\ee
where $N$ is the number of particles along the boundaries of the square.

Note that it is not obvious if the solutions of the KPI equation 
will also be a solution of the lattice model.
However, it was shown in \cite{IPA} that this approach has been successfully 
considered in a  one-dimensional lattice model.
Specifically, when a one-dimensional lattice model with long-range
interactions was considered (which, in the continuum, keeps
its nonlocal behaviour) the long time evolution of the 
localized waves is governed by an asymptotic integrable equation 
 the so-called Benjamin-Ono equation which allows the explicit 
construction of moving kinks on the lattice.
Accordingly, in this section we investigate the dynamical behaviour 
of the lattice model (\ref{latt}) numerically, using
as initial conditions the one and two soliton solution of the
 KPI integrable equation given in section 3.2. 

By comparing the coefficients and the variables of equations (\ref{KPa})  and 
(\ref{KP}), it is easy to see that, (\ref{KPa}) transforms to (\ref{KP})
when
\bea
\bar{\alpha}=3, &\hs \hs& x \ra \xi=x-ct\nonumber\\
\beta=1, &\hs \hs& y \ra Y=\ve^{1/2}y\nonumber\\
\bar{\delta}_L=1, &\hs \hs &t \ra  T=-\ve t/2c.
\label{cond}
\eea
Then it is a matter of simple algebra to observe that, 
due to (\ref{cond}),
 the parameters of the  discrete model given by 
(\ref{del}-\ref{epar}) are equal to
\be
\alpha_2=3\ve, \hs \hs \delta=\fr{c^2}{12}-16\eta-\ve
\ee
and, thus, the only arbitrary parameters of the lattice model are
the following three:  $\ve$, $c$ and $\eta$.
Due to numerical simplicity, in all our simulations we have chosen the
following fixed values for:
(i) the small parameter  related with the weak nonlinearity  
$\ve=0.1$,  (ii) the lattice spacing $h=1$,  
(iii) the total number of the lattice points $N=100$, 
(vi) the time step $dt=0.1$ and (vi) the 
total number of time steps equal to 5000 points.
However, different values for the above parameters give qualitatively
the same results as long as $\ve<1$.\\

$\bullet$ {\it Dynamics of  One Soliton}\\

\begin{figure}
\vskip -5cm
\hskip 2cm
\put(100,190){$t=0$} 
\hskip 5cm
\epsfxsize=10cm\epsfysize=10cm\epsffile{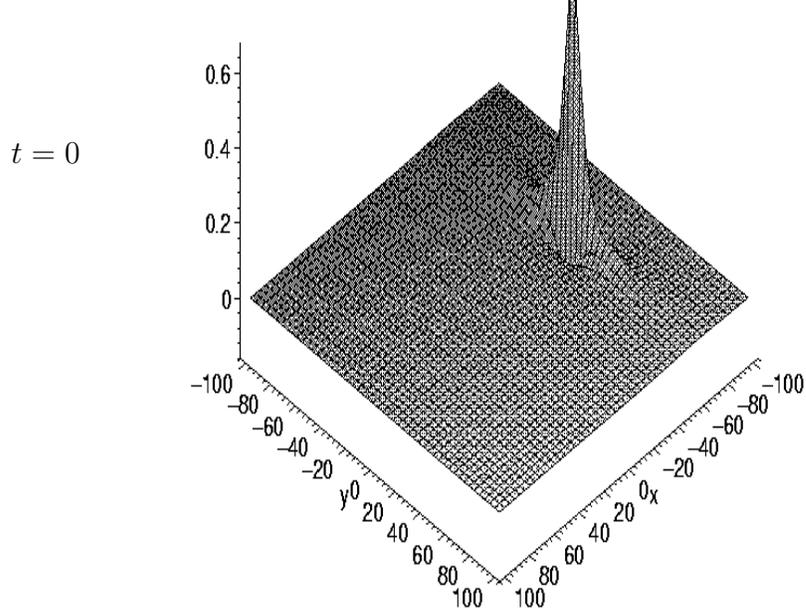}
\caption{The one soliton solution (\ref{l1sol}) for $\ve=0.1$.}
\label{fig-1s}
\end{figure}

Firstly, we investigate the time evolution of a single solution
where the configuration given by (\ref{1so}) have been considered
 under the change of variables (\ref{cond}).
More precisely, the initial conditions for the 
displacements and the velocities of each lattice particle, 
in the case of the one soliton (\ref{1so}), are given by 
the analytic expression
\be
u(x,y,t){\big |}_{t=0}=4\fr{-\left(x-ct+\ve t/2c+60\right)^2+\ve y^2+3\,}
{\left[(x-ct+\ve t/2c+60)^2+\ve y^2+3\right]^2}{\biggl |}_{t=0}
\label{l1sol}
\ee
and its time derivative, respectively. 
Figure \ref{fig-1s} illustrates snapshot of the displacement $u_{i,j}(t)$
at $t=0$.

We have run our simulations for different values of 
the parameters $(c,\eta)$ and investigate whether the initial 
lattice soliton  relaxes and propagates in  the lattice space 
with small (or no) oscillations.
This phenomenon occurs, only, for specific range of values of the 
aforementioned parameters which give qualitatively the same results.
In particular, due to \cite{P4}, we know that the noncentral interaction 
parameter have to be negative and small.
Our numerical study have shown that for values of the parameter 
$\eta<1/60$ and when the soliton velocity  takes values between
$1\le c \le 4$ the loss of energy by radiation of small 
amplitude waves is small.

\begin{figure}
\vskip -2cm
\hskip 4cm
\epsfxsize=19cm\epsfysize=19cm\epsffile{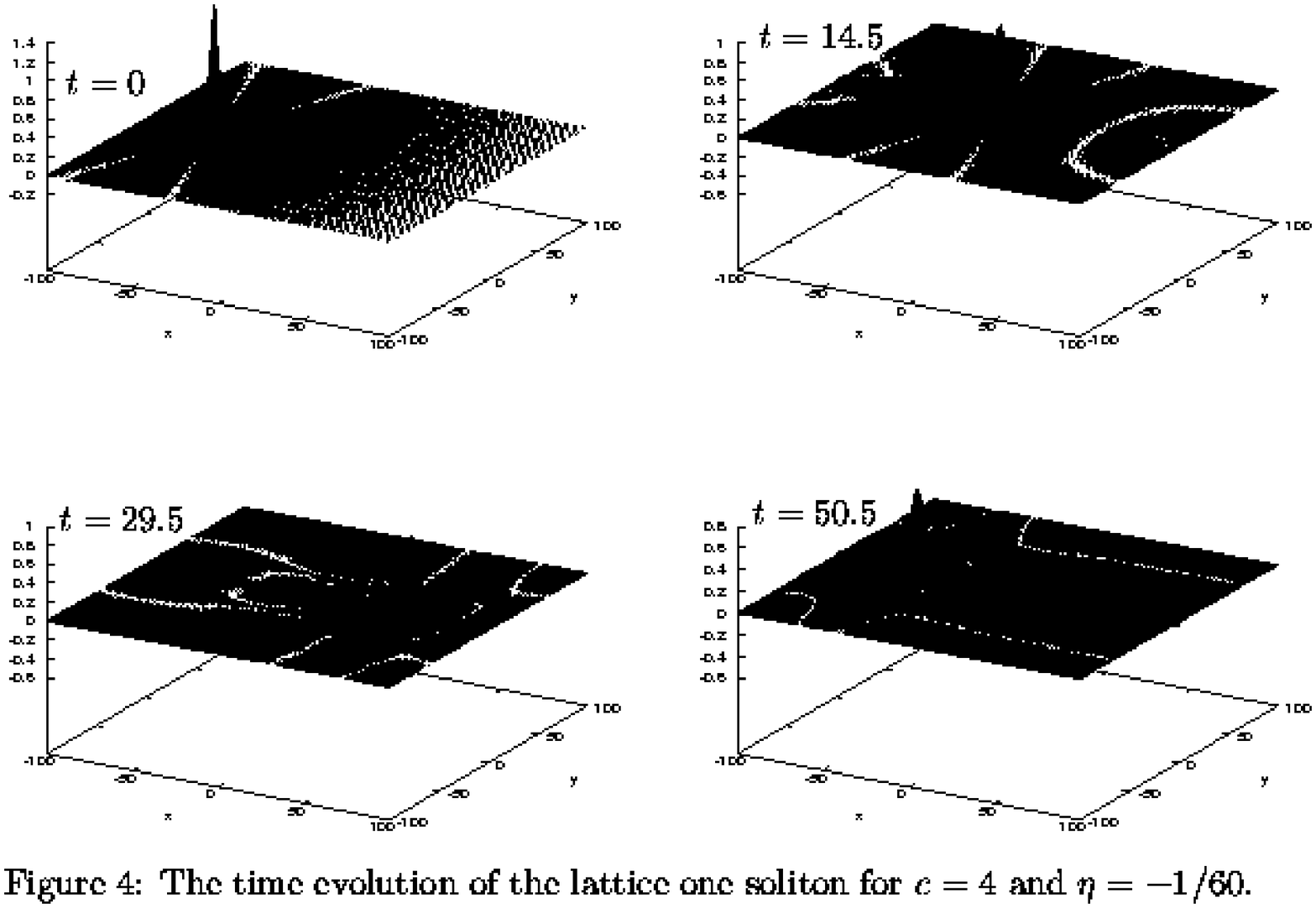}
\label{fig-1sn}
\end{figure}

In figure \ref{fig-1sn}, we display a full three-dimensional plot
corresponding to the displacement $u_{i,j}$ at four different times
$t=0, 14.5, 29.5, 50.5$.
The initial conditions were created from one soliton placed  at the
position $x_0=-60$. 
 Figure \ref{fig-1sn} shows that the lattice soliton
 is stable throughout the  numerical simulations; 
more precisely,  its size is constant as it moves
towards the plane without emitting any radiation.
This process continues through several cycles,
due to the periodic boundary conditions,
without the occurrence of  instabilities since the 
configuration settles to a single soliton which move
 along the lattice grid.

Qualitatively, similar results have been found and
studied in great detail in $\cite{IPA}$ for one-dimensional
lattice kinks which are solutions of a
nonlocal  discrete model with  long-range interactions. 
In that case, the lattice radiation was inversly proportional 
 to the kink thickness.
\\

$\bullet$ {\it Two Soliton Scattering}\\

Next, we discuss the results of a numerical evolution of the full 
time-dependent lattice equations (\ref{latt}), in order to investigate the
interaction and scattering of two solitons.

We take two solitons initially located at the origin
 (represented in Figure \ref{fig-2s})
and evolve the equations of motion for different values 
of the soliton velocity $c$ and of the noncentral force parameter $\eta$.
For values (approximately) between the following range:
 $4\le c \le 8$ and $-1/30 \le \eta \le -1/360$
the numerical simulations  give qualitatively the same results, 
represented in Figure 5.

From Figure 5, we see that the two solitons which initially
(at $t=0$) form 
a single large soliton travel with opposite velocities away 
from each other and then, due to the periodic boundary conditions, 
 reappear at the edge of the lattice.
Now, the two solitons head towards each other until they
merge to form a single soliton again (at $t=48.5$) 
while after the collision, they continue their path 
with no phase shift or emission of radiation, ie they
scatter trivially as the KPI solitons.
In fact a small amount of energy is radiated during the soliton motion
as a result of the highly nonlinear deformation the solitons suffer
in adjusting to the lattice grid.
This radiation can be dealted with numerically by applying absorbing 
boundary conditions at the edge of the grid.
The radiation emission is eliminated as the soliton velocity 
$c$ increases.
This scattering behaviour is continuous throughout for a long time;
specifically, we have run our simulations up to $t_f=500$ and although
the lattice grid is covered from radiation the two solitons
are easily distinguished.

\begin{figure}
\vskip -2cm
\hskip 4cm
\epsfxsize=19cm\epsfysize=19cm\epsffile{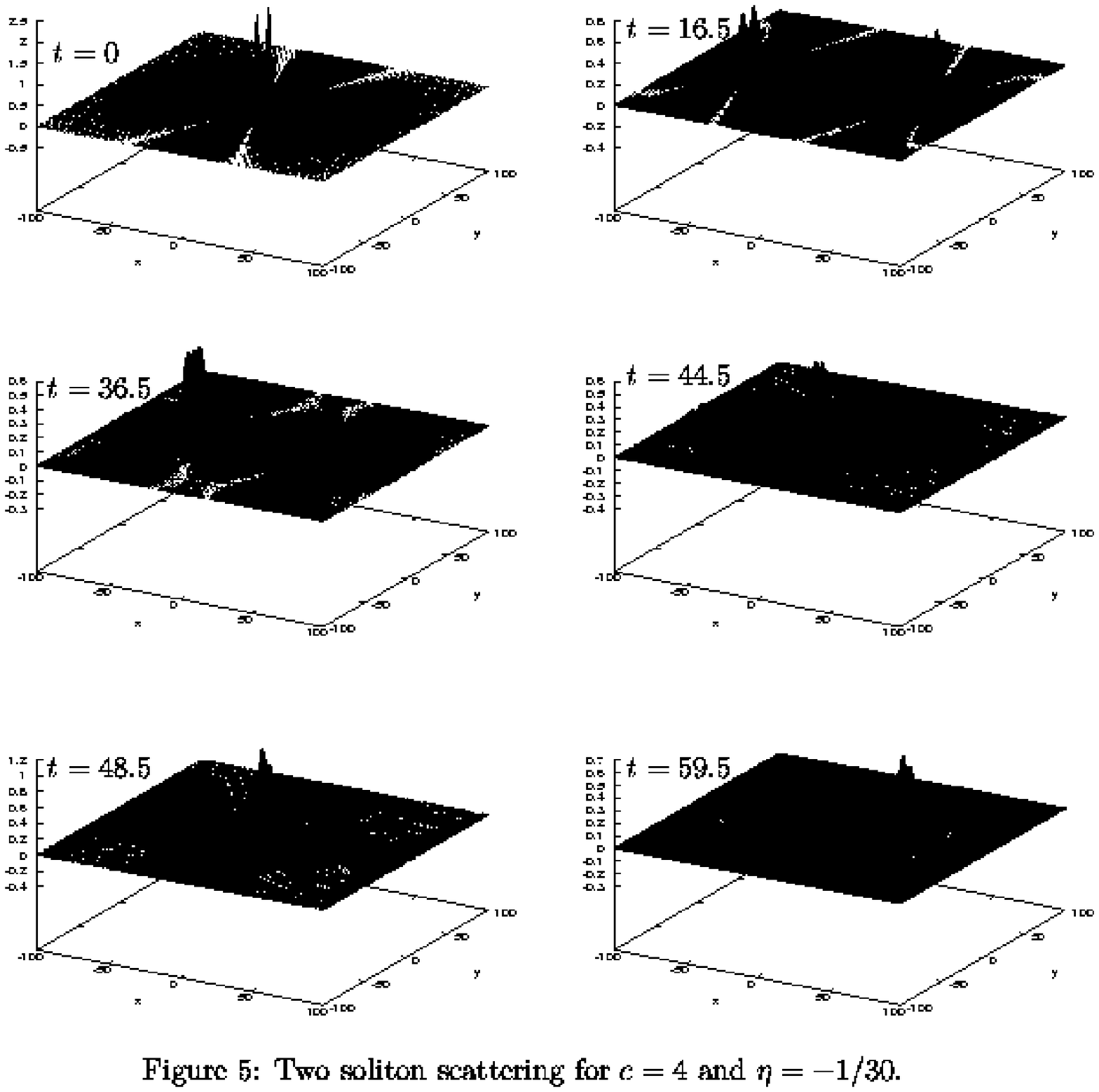}
\end{figure}

To conclude, this approach can be extended to higher soliton numbers and
we expect the results to be qualitatively like the ones obtained
in the first two cases of the one and two soliton dynamics.

\section{Conlusions}

Tha main objective of the present work is to examine the long-time 
evolution of nonlinear localised waves of soliton type propagating on a 
two-dimensional lattice model.
The starting lattice model involves nonlinear (nonconvex lattice 
potential) and competing interactions which follow from noncentral 
interactions and play a key role in the existence and stability of the 
localised waves.
Physically, the model  describes the mechanism of 
micro-twinning of small ferroelastic domains in alloys suffering
phase transformations like ferroelastic materials, martensitic 
transformations in shape memory alloys, etc.

The most interesting point of the study is that, on the basis of a 
multi-scale  technique, an asymptotic model for the long-time evolution of 
the nonlinear waves has been derived.
This model is then governed by Kadomtsev-Pietvishviali I equation which
possesses soliton solutions. 
In order to check the analytical conjecture given by the asymptotic model, 
some numerical simulations have been performed (directly) on the discrete 
system.
In fact, the numerical results describe the time-evolution of 
one- and  two-soliton solutions.
It appears that the soliton solutions thus generated are particular robust 
and stable which implies that the prediction provided by the 
asymptotic model is a good approximation of the solution of the 
discrete system.
Therefore, we conclude with the observation  that the KPI equation is an 
efficient asymptotic  model to predict the existence of localised objects 
or nonlinear  excitations of soliton type in a rather complex physical 
system.

Further extensions of the present work can be investigated. 
In particular, the dynamics of localised waves for a lattice model 
including higher order nonlinear interatomic terms in the potential 
(fourth order  nonlinear term in discrete deformation, see equation 
(\ref{poten})) can be  considered; while it is  worth investigating the 
influence of applied  forces and damping on  the nonlinear wave 
propagation and stability.

\section*{Acknowledgements}
This work has been performed in the framework of the TMR European Contract 
{\bf FMRX-CT-960062:} {\it Spatio-temporal instabilities in
deformation and fracture mechanics, material science and nonlinear physics
aspects}.
TI acknowledges the Nuffield Foundation for a newly appointed lecturer award.

\end{document}